# Generative AI Perceptions: A Survey to Measure the Perceptions of Faculty, Staff, and Students on Generative AI Tools in Academia

*Sara Amani, Lance White, Trini Balart, Laksha Arora, Dr. Kristi J. Shryock, Dr. Kelly Brumbelow, and Dr. Karan L. Watson*

## Abstract
ChatGPT is a natural language processing tool that can engage in human-like conversations and generate coherent and contextually relevant responses to various prompts. ChatGPT is capable of understanding natural text that is input by a user and generating appropriate responses in various forms. This tool represents a major step in how humans are interacting with technology. This paper specifically focuses on how ChatGPT is revolutionizing the realm of engineering education and the relationship between technology, students, and faculty and staff. Because this tool is quickly changing and improving with the potential for even greater future capability, it is a critical time to collect pertinent data. A survey was created to measure the effects of ChatGPT on students, faculty, and staff. This survey is shared as a Texas A&M University technical report to allow other universities and entities to use this survey and measure the effects elsewhere.

## I. Introduction
ChatGPT created by OpenAI, along with other generative artificial intelligence (AI) systems that have been released in the late 2022 and early 2023 have had extensive media coverage focusing on the power of these AI systems. Speculation about the potential of AI systems given the impressive performance of these released systems has been widespread with some media outlets concerned that generative AI may be more proficient in things like creative art than most people. ChatGPT is of particular interest for this work due to the text-based nature of the system. The training data used for the GPT-3 based ChatGPT system spans 570 GB of text data available on the internet up to the month of September 2021.

### *A. Intended Survey Audience*
To understand the perceptions of generative AI at Texas A&M University in College Station, two surveys were developed and distributed to the students and faculty/staff on campus. While some questions were mirrored between the populations about gathering feedback and experiences of AI systems, the first survey sent to students focused on their perceptions of ChatGPT and how they were using it for their academic study. The second survey shared with faculty and staff focused more on their thoughts regarding how generative AI is affecting their courses and how they think it should be used by students in their classes. The surveys were distributed through the bulk mail system to all students, faculty, and staff at the university with targeted emails sent to faculty and staff in the college of engineering.

### *B. Intended Survey Use*
Results from the survey are intended to inform the educational community on perceptions, misconceptions, concerns, and insight into the use of generative AI systems in higher education. Universities can use this knowledge to respond and reflect in real time with the goal of improving the academic environment. We expect this survey will provide knowledge that can improve the current and future trends in education.

***C. Survey Access***

We have included the survey questions used in the appendix below. The original survey was created using Qualtrics with a unique link created for both the student and faculty/staff surveys. Please reach out to us if we can assist you with gathering feedback from your institution. We are happy to share a copy of the survey, so that you may simply edit the questions for your own needs if it is helpful. Feel free to email one of the authors for access.

***D. Survey Topics***

The survey topics were decided in order to cover a wide range of areas related to the use of ChatGPT in academia and its potential impact on education.

*1. Uses of ChatGPT (*Baidoo-Anu & Owusu, 2023)*:* How ChatGPT is being used in academia, including areas such as content generation, personalized learning, research assistance, tutoring, and assessment. The perceived benefits and limitations of using ChatGPT in these contexts, as well as any concerns or challenges associated with its use.

*2. Comfortability with using ChatGPT in academia:* Level of comfort among the community with using ChatGPT in academic settings. Factors that may influence comfort level, such as familiarity with AI technology, prior experience with ChatGPT, and attitudes towards technology integration in education.

*3. Academic dishonesty:* Perceptions and experiences related to academic dishonesty facilitated by ChatGPT. Attitudes towards cheating and plagiarism using ChatGPT, awareness of potential risks and consequences, and strategies to prevent academic dishonesty.

*4. Utilization of external resources for course aid:* Extent to which students and educators use external resources, including ChatGPT, for course aid. Types of resources used, reasons for using them, and perceptions of their impact on learning outcomes.

*5. Competencies for engineering:* The perception of the impact that ChatGPT would have on different essential skills for engineers and important aspects of engineering education and student development.

Some of the essential skills evaluated (Male, Bush, & Chapman, 2011):

- Critical Thinking: Engineers must be able to analyze and evaluate problems in a systematic and logical way.
- Problem Solving: Engineers must be able to develop innovative solutions to complex problems.
- Teamwork: Engineers often work in teams, and they need to be able to collaborate effectively with colleagues from different disciplines.

In addition to the essential skills for engineers, the investigation explores other important aspects of engineering education and student development. These include (Stefanou & Salisbury-Glennon, 2002):

- Self-efficacy: The belief in one's ability to perform tasks and achieve goals. It is an important factor in engineering education and can impact students' academic and career success.
- Test Anxiety: The feelings of unease and apprehension that some students experience before, during, or after tests. Test anxiety can negatively impact academic performance and may affect student motivation and engagement.
- Academic Performance: The measure of a student's achievement in a particular course or program. Academic performance is an important factor in engineering education, and it is influenced by a range of factors, including the students' motivation, engagement, and study habits.
- Intrinsic Motivation: The drive to engage in an activity for its own sake rather than for external rewards. Intrinsic motivation is an important factor in student engagement and academic achievement.
- Student Engagement: The degree to which students are involved in and committed to their learning. Engaged students are more likely to be motivated, perform better academically, and have higher levels of satisfaction with their education.

This topic explores the role of ChatGPT in developing or enhancing these skills, and attitudes towards the integration of ChatGPT in engineering education.

*6. How AI will impact the future of various disciplines:* Perceptions and expectations of how AI will impact different disciplines, such as science, technology, engineering, arts, mathematics, social sciences, humanities, and beyond. Anticipated benefits, challenges, and implications of AI in various fields of study.

### *E. Survey Topics Addressed Qualitatively and Quantitatively*

Based on the survey topics listed above, the survey refers to several quantitative and qualitative questions targeting faculty, staff, and students. Regarding the uses of ChatGPT, both surveys asked “In what ways have you used ChatGPT?” as a quantitative question where students could select from a range of uses, which differed slightly based on the two surveys. In the student survey example some potential choices included “asking technical questions”, “solving homework”, and “writing essays”, which in the faculty/staff survey we asked about “grading writing assignments”, “writing technical documents” and “asking general knowledge questions”, among others. As a qualitative measure, if something wasn’t listed, an “Other” option with a text box was included. We also quantitatively asked separate questions targeting academic dishonesty. In the faculty and staff survey we asked “How likely do you think students are to violate the Aggie Honor Code now that ChatGPT is released to the public?”. In the student survey we asked “To what extent is using ChatGPT ethical/appropriate for coursework?”. These questions are referencing the Aggie Code of Honor, a series of statements that are in place to hold the students of Texas A&M University to a high standard of ethics and dignity. As a qualitative question, in both surveys we asked “How do you think AI tools like ChatGPT will impact the future of your discipline?” which targeted the survey topic of how AI will impact the future of various disciplines.

## II. Preliminary Findings and Guidance

Preliminary analysis of survey results provides interesting and useful insights related to generative AI systems.

### *A. Survey Responses*

In four weeks of survey availability, we have had 243 faculty and staff (101 faculty and 142 staff) and 813 students complete the survey. For the faculty and staff, 101 respondents have been people whose primary responsibility is as a faculty member. A total of 142 respondents are people classified as academic staff, most of whom have major responsibilities requiring interfacing with students as advisors or instructors for non-credit freshmen success courses. For both the faculty/staff and student surveys, the largest college represented in the respondents is engineering. Out of 243, there are 189 faculty and staff from the college of engineering with the next two largest numbers of replies being the College of Arts and Sciences and College of Agriculture and Life Sciences with six respondents each. Out of 813 students (representing both undergraduate and graduate students), there are 647 students from the College of Engineering, 30 from the College of Arts and Sciences, and 28 from the School of Education and Human Development comprising the largest populations represented in the sample.

### *B. Comments from Faculty and Staff*

There were several open-ended question types asked in the survey. We have highlighted some of the themes and direct quotes, in no particular order, received from faculty and staff who completed the survey.

1. There is a clear tendency to see artificial intelligence as something that is here to stay and that we will have to learn to live with as a society.
    - “AI tools like ChatGPT are here to stay, so I'm on the "if you can't beat 'em, join 'em" boat!”
    - “Generative language tools (like ChatGPT) will increasingly become part of our everyday lives”
    - “AI technology is not going away”

2. The importance of adapting the way we are teaching and evaluating students in order to be able to see artificial intelligence as a tool and not as a threat is highlighted.
    - “In terms of teaching, it will force us to evaluate students' learning differently”
    - “They will challenge institutions of learning to find better ways to assess the value of the education being offered.”

3. Risks are mentioned repeatedly as well with concern about students understanding the limitations of artificial intelligence and using it judiciously. Emphasis should be placed on these tools that do not always have the right answer and enhance critical thinking on the part of students.

- "I worry that students will use it without understanding its limitations."

4. The conditionality of the impact of this tool, positive or extremely positive if and only if used correctly, is repeatedly stressed.
    - "ChatGPT is a powerful tool. If we use it properly and wisely, it will leverage personal knowledge tremendously. Look at the positive side of it. Every revolution takes time to realize, and one should adapt themselves to it."
    - "I think if we use it effectively it could be very helpful for learning. It can help move us into a more technology focused future."

5. On the other hand, there are multiple opinions about the negative impact that this tool will have, diminishing the reasoning abilities and critical thinking of students. There are those who consider ChatGpt only as a tool for cheating on exams.
    - "Enables self-learning but simultaneously reduces critical thinking."
    - "I think students' critical writing skills will diminish further."
    - "I am definitely worried of an overreliance on ChatGPT where students abuse the knowledge to get answers without putting in any effort themselves first"

***C. Comments from Students***

We have highlighted some of the themes and direct quotes, in no particular order, received from students as well who completed the survey.

1. Among students, it appears that many have a positive view of the use of AI tools, such as ChatGPT in education. They highlight the advantages of having access to an AI tutor who can help understand concepts, offer examples, review papers, provide study tips and mental health resources, and collaborate on personal projects. The ability of AI to streamline information and provide quick summaries also appears to be highly valued. There is a belief that AI tools can be very beneficial in enhancing learning and research, but they also stress the importance of responsible use.
    - "I have adopted ChatGPT as my own personal tutor. When I struggle understanding a concept, it can help rephrase a concept and provide me with examples."
    - "I personally use it as a tool to explain concepts that I would be searching for, as it gives a more streamlined answer and explains it in simple terms."
    - "I believe the future of my discipline will be impacted in a positive manner. For example, the ability to summarize code in an instant saves a lot of time. ChatGPT also provides the benefit of not needing to sift through outdated and/or poorly written documentation."
    - "I personally see the positive side of the AI tools in enabling us to speedily research things and get a better idea of how to approach problems."

2. On the other hand, students also recognize the potential for misuse and plagiarism, and suggest that it is up to the individual to use the tool ethically and as an aid to learning, not as a substitute for their own work. There is concern that ChatGPT may lead students to take shortcuts, cheat, and not really understand the material. In addition, some are concerned that AI may reduce the need for students to develop essential skills such as critical thinking, creativity, and interpersonal communication. A number of potential drawbacks are mentioned that need to be carefully considered.

- “How useful is a degree if the students who earn them do so by cheating the system? ChatGPT gives students the opportunity to cut corners and take the "easy way out."
- “Of course, there are also negatives one must consider. For example, instead of coming up with creative or thoughtful responses, I could instead ask ChatGPT to write points on how it negatively affects discipline in fields: critical thinking, diminished expertise, reduced creativity…, ethical concerns, reduced interpersonal skills, misinformation.”
- “I am concerned for how AI will reduce the need for engineers or those studying the engineering discipline to recognize the value of developing skills related to knowledge synthesis.”

3. Some students express a neutral or balanced perspective on ethical issues arising from the use of ChatGPT. Potential benefits and drawbacks of the technology are acknowledged, and opinions about its use vary depending on context and individual responsibility. While some believe that AI can be a valuable reference tool if used ethically and with integrity, others caution against the loss of critical thinking skills and creativity that could result from over-reliance on AI. It is recognized that ethical use of AI depends on the intentions and actions of the user, and that education and incentivizing integrity could be more effective than strict policing of cheating methods.
    - “It can be used negatively but that depends on the student and their intended use.”
    - “I think if we learn how to use it in an ethical, helpful way, we can use it to our advantage.”
    - “I feel like it's very dependent on certain conditions, just like this survey. There's many different views and standpoints you could take on the ethical issues arising from ChatGPT. I think it's an incredible technology and very effective at providing information, however, it is also possible to be used to solve homework and give answers which is academically wrong.”
    - “Just like any tool, I personally believe that AI can be used for both great good and bad.”

4. Some students related ChatGPT to other already existing tools with which higher education must reckon.
    - “I think educators need to learn how to HELP students use it in a way that promotes critical thinking and problem solving. Students should learn how to use it as a tool (like when we incorporated the calculator!) rather than a tool for cheating. Instructors could have students diagnose ChatGPT answers and critique them for accuracy or how they might elaborate on the response of ChatGPT.”

***D. Highlights of Data Received***

The survey includes several questions related to the use of ChatGPT and the implications on dishonest behaviors and academic success. We have provided some highlights and comparisons between the faculty/staff and students who completed the survey, in no particular order of importance.

1. Related to use of ChatGPT, 64% of faculty/staff and 73% of students have an account and have used it. The percentages of the faculty and of staff who have used it were similar.

2. When asked how they have used ChatGPT, three responses were received from faculty and staff much more than the other options. These include: ask technical questions (20%), ask general knowledge questions (18%), carry on a conversation out of curiosity (16%).

   For students, the same three responses for using ChatGPT were also in their top four, but they also listed an additional area within the highest replies. These include: ask technical questions (17%), explain concepts (16%), ask general knowledge questions/advice (16%), and carry on a conversation out of curiosity (15%).

   When students were asked how their peers use ChatGPT compared to how they use it, however, responses such as solving homework and writing essays more than doubled. Five percent of students reported using it to solve homework, but over 11% replied that their peers had used it for this purpose. Similarly, 4% of students responded they use ChatGPT for writing essays but 11% felt their peers had used it for this reason.

3. When faculty and staff were asked how comfortable they felt with their students using ChatGPT in their courses, 47% felt somewhat comfortable or extremely comfortable with it with an additional 18% being neither comfortable nor uncomfortable.

4. Related to academic dishonesty, responses from faculty/staff definitely shifted from feeling less likely that their students would violate the Aggie Honor Code prior to the rollout of ChatGPT to being more likely they would violate it now that ChatGPT has been released to the public. Prior to the rollout of ChatGPT, 35% listed it as somewhat likely or extremely likely students would be dishonest with 30% responding neither likely nor unlikely. After the rollout though, 55% replied as somewhat likely or extremely likely students would engage in academic dishonesty with similarly 31% feeling neither likely nor unlikely.

   Similarly, when students were asked if they felt ChatGPT would enable academically dishonest behaviors, 63% replied that they somewhat agree or strongly agree. Sixteen percent selected they neither agree nor disagree with that statement.

5. Faculty and staff were surveyed on which uses of ChatGPT they felt would be most beneficial for students, and they were allowed to select all of the options they felt applied. The top two benefits for students they selected were personalized learning (27%) and effective and instant feedback (25%). The remaining options received less than 15% responses.

6. Faculty, staff, and students were asked how ChatGPT will impact the following: critical thinking, problem solving, teamwork, self-efficacy, text anxiety, academic performance, intrinsic motivation, and student engagement. There were many similarities in the replies between the two groups of respondents. The areas both faculty/staff and students identified as ChatGPT most negatively impacting were critical thinking and problem solving. Students replied much more positively that academic performance would be helped with ChatGPT than faculty and staff did.

### *E. Visual Representations*

To provide a visual representation of the feedback received from faculty, staff, and students in response to the open-ended question related to how AI tools like ChatGPT will impact the future of their discipline, a word cloud was created for each population, which are depicted in Figures 1-3.

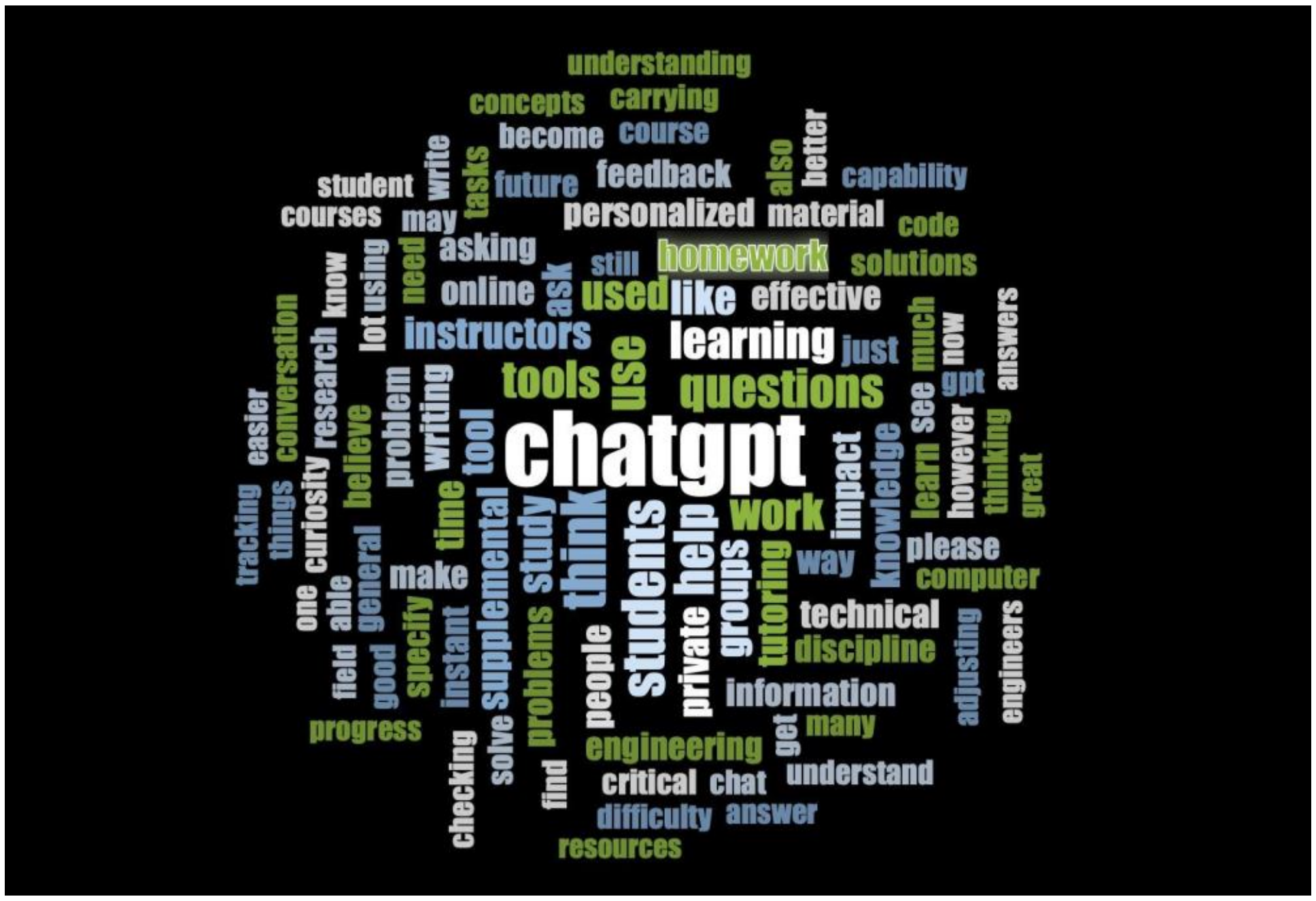

Figure 1. Word cloud created based on only faculty responses to how ChatGPT will impact the future of their discipline.

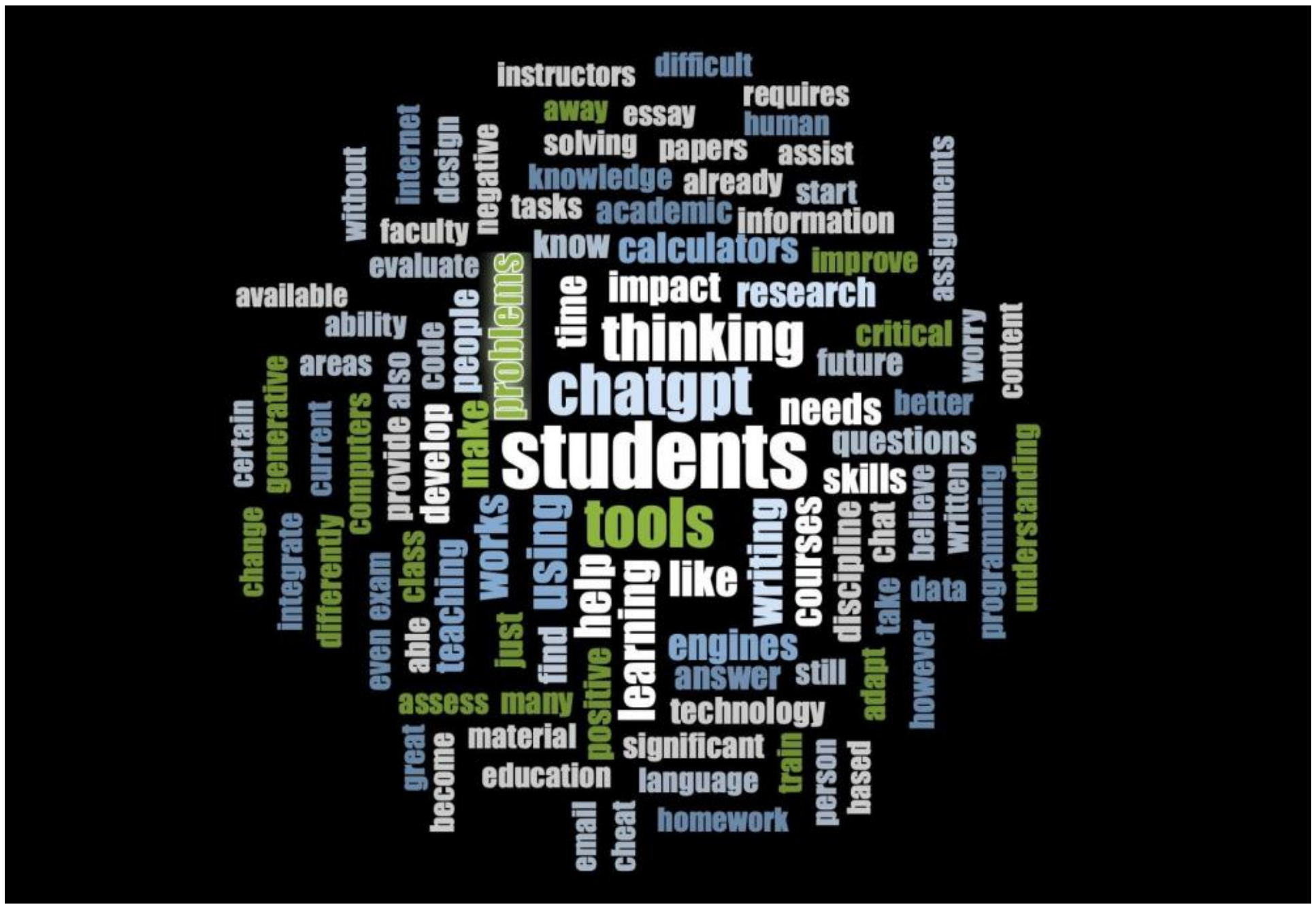

Figure 2. Word cloud created based on faculty and staff responses to how ChatGPT will impact the future of their discipline.

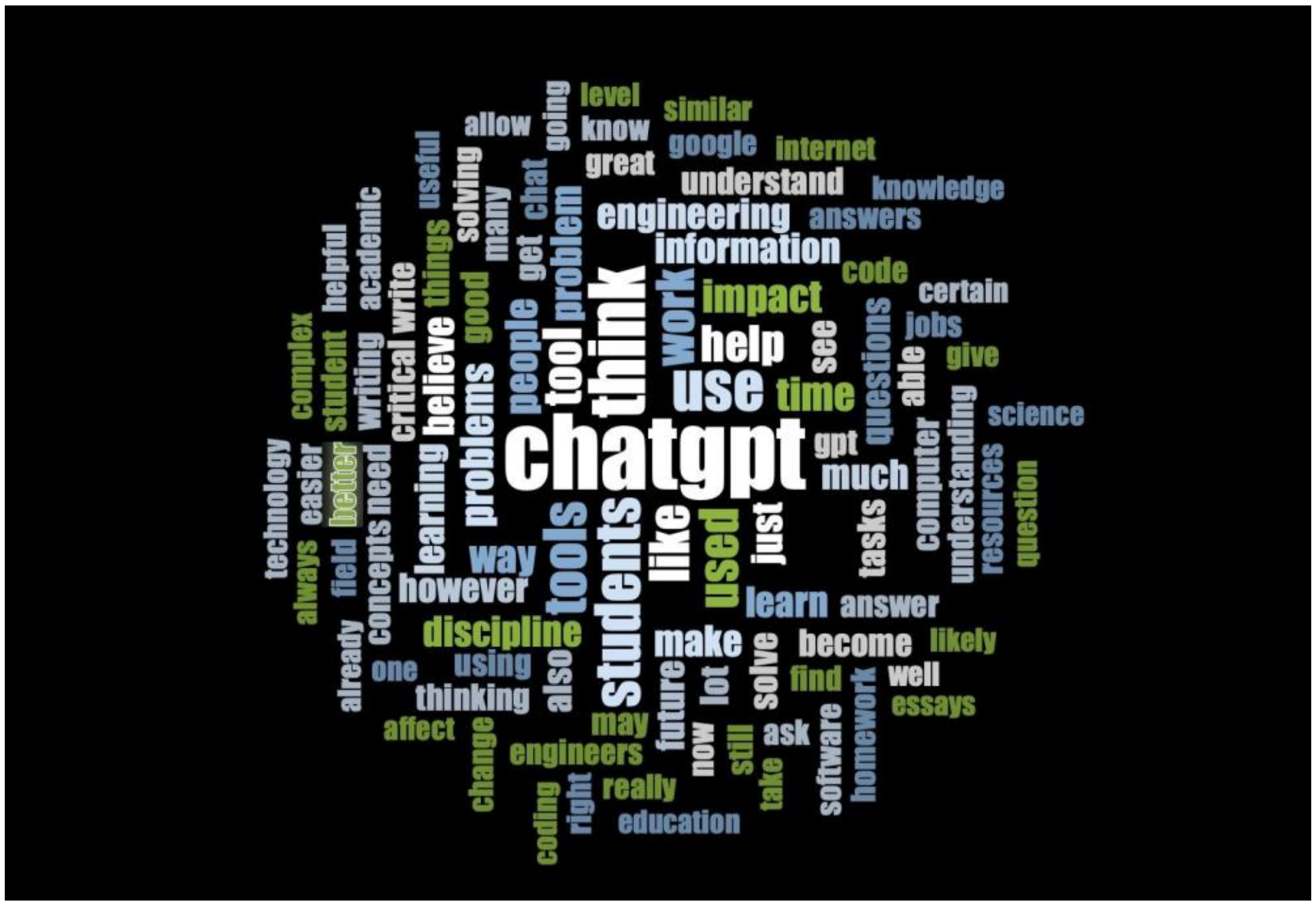

Figure 3. Word cloud created based on student responses to how ChatGPT will impact the future of their discipline.

## III. Recommendations for Future Survey Distribution

After distribution of the survey, the authors have provided a few details that might be considered in the event of future survey iterations by interested parties.

1. Incorporate a dropdown box when asking for estimated graduation date by semesters.
2. Use a calendar tool or a standard date format inquiring about the date of account creation.
3. Decide on if having a response provided for each of the questions is important. (Each of the questions in our implementation were not required to have a response.)
4. Revise college and major dropdown choices to be based on the individual institution using this survey.
5. Consider opening the staff survey to include staff who are outside of the academic units of the university system, such as IT and technical staff who are using ChatGPT for their professional job duties.
6. Establish how staff members are involved in the In the interest of inclusivity at this institution, staff are able to teach in some capacities that may not be conventional at other institutions.

## IV. Acknowledgements

The authors want to acknowledge the support of the faculty, staff, and students of Texas A&M University at College Station, TX, for completing this survey and providing insightful feedback. In addition, we appreciate the support of the College of Engineering for distributing the surveys to the faculty and staff in the college for distribution.

## V. About the Authors

Sara Amani

Sara Amani is a doctoral student at Texas A&M University. She completed her Bachelors of Science in chemical engineering from Texas A&M University at Qatar. She is currently pursuing her PhD in Multidisciplinary Engineering with a focus in engineering education. Her research interests include mental health in engineering education and women in engineering.

Lance L.A White

Lance L.A. White is a Ph.D. candidate at Texas A&M University in Interdisciplinary Engineering housed within the Multidisciplinary Engineering department with a thrust in Engineering Education. He is working as a graduate research assistant at the Institute of Engineering Education and Innovation at the Texas Engineering Experiment Station. His research centers on diversity, equity, inclusion, and access in the context of Engineering in higher education. His dissertation work is looking at Engineering degree program curricula to understand impacts of institution types and commitment to servingness of underrepresented populations in engineering. He is trained in both qualitative and quantitative research and plans to pursue a tenure-track faculty position in the near future. His dissertation chair is Dr. Karan Watson, and he is working under the Director of IEEI, Dr. Tracy Hammond.

Trini Balart

Laksha Arora

Laksha Arora is a freshman undergraduate engineering student at Texas A&M University, and is interested in pursuing a career in computer science. Additionally, Laksha has an interest in artificial intelligence, machine learning, and cybersecurity. Outside the lab, she is an active member of the Aggie Women in Computer Science Club and Aggie Coding Club.

Dr. Kristi J. Shryock

Kristi J. Shryock, Ph.D., is the Frank and Jean Raymond Foundation Inc. Endowed Associate Professor in Multidisciplinary Engineering and Affiliated Faculty in Aerospace Engineering at Texas A&M University. She also serves as Director of the Novel Unconventional Aerospace Applications iN Core Educational Disciplines ($NUA^2NCED$) Lab and of the Craig and Galen Brown Engineering Honors Program and National Academy of Engineering Grand Challenges Scholars Program. She has made extensive contributions to the methodology of forming the engineer of the future through her work in creating strategies to recruit, retain, and graduate engineering students. The network of transformational strategies she has developed addresses informing early, preparation for success, increasing diversity of the field, establishing strong identity as an engineer, and enhancing critical thinking and professional skills.

Dr. Kelly Brumbelow

Kelly Brumbelow, Ph.D., P.E., has been an engineering faculty member at Texas A&M University since 2002. He is the Associate Head of the Department of Multidisciplinary Engineering and is an associate professor in that department as well as the Zachry Department of Civil and Environmental Engineering. His research focuses on engineering education, particularly looking at multi- and inter-disciplinary programs, water resources planning and management in many contexts, and efficiency and security of water distribution systems.

Dr. Karan L. Watson

Karan L. Watson, Ph.D., P.E., is currently Provost Emeritus and a Regents Senior Professor of Electrical and Computer Engineering, having joined the faculty at Texas A&M University in 1983 as an Assistant Professor. She served as the Co-Director of the Institute for Engineering Education and Innovation and is currently a distinguished fellow of this Institute. She has served in numerous administrative roles at Texas A&M University, including, provost and executive vice president, vice provost, dean of faculties and associate provost, interim VP for diversity, associate dean of Engineering, and program chair for interdisciplinary engineering. Dr. Watson is a fellow of three organizations The Institute of Electrical and Electronic Engineers (IEEE), the American Society for Engineering Education, and the Accreditation Board for Engineering and Technology (ABET). Her awards and recognitions include the U.S. President's Award for Mentoring Minorities and Women in Science and Technology, the American Association for the Advancement of Science mentoring award, the IEEE International Undergraduate Teaching Medal, the American Society for Engineering Education Lifetime Achievement

Award, and numerous faculty awards at Texas A&M University. She has served as President of the Accreditation Board for Engineering and Technology (ABET) and the President of the Education Society of IEEE.

## VIII. Surveys

**A. ChatGPT Faculty Survey**

Dear TAMU faculty /staff,

A team of researchers led by Dr. Kristi Shryock are interested in understanding how ubiquitous the AI platform ChatGPT is in the lives of our faculty / staff. This short survey is seeking to gauge the use cases of ChatGPT among faculty / staff and understand the prominence of ChatGPT among our faculty / staff at Texas A&M University. All responses to this survey will be anonymous.

Please indicate whether you are employed as faculty or staff at Texas A&M University

o Staff

o Faculty

(If faculty) Please indicate your position or status at the university

Track

Classification

▼ Faculty ... Faculty ~ Tenure Track ~ Professor

(If faculty) Please select the college/school and department/program you are associated with

College/School

Department/Program

▼ College of Agriculture and Life Sciences ... Transition Academic Programs ~ General Studies

How familiar are you with ChatGPT by OpenAI?

o Extremely familiar

o Very familiar

o Moderately familiar

o Slightly familiar
o Not familiar at all

Have you made an account and used ChatGPT for any reason (personal or educational)?
o No
o Yes

(If yes to account) When did you decide to create your account? (month or semester you began using ChatGPT)

(If yes to account) In what ways have you used ChatGPT? (select all that apply)
▢ Asking technical questions
▢ Carrying on a conversation out of curiosity
▢ Ask general knowledge questions
▢ Preparing materials for your courses
▢ Grading writing assignments
▢ Checking for solutions capability of ChatGPT for your course work
▢ Writing technical documents
▢ Querying ChatGPT for research related activities
▢ Have not used
▢ Other (Please Specify)

How comfortable would you be with your students using ChatGPT in your courses?
o Extremely comfortable
o Somewhat comfortable
o Neither comfortable nor uncomfortable
o Somewhat uncomfortable
o Extremely uncomfortable

How likely do you think your students were to violate the Aggie Honor Code prior to the rollout of ChatGPT?
o Extremely likely
o Somewhat likely
o Neither likely nor unlikely
o Somewhat unlikely
o Extremely unlikely

How likely do you think your students are to violate the Aggie Honor Code now that ChatGPT is released to the public?
o Extremely likely
o Somewhat likely
o Neither likely nor unlikely
o Somewhat unlikely
o Extremely unlikely

How much do you agree or disagree with this statement: "Students should be allowed to utilize resources not provided by the instructor of a course." (Ex: Chegg, Coursehero, ChatGPT, Quizlet)
o Strongly agree
o Somewhat agree
o Neither agree nor disagree
o Somewhat disagree
o Strongly disagree

What resources are acceptable for students to use outside of those provided by an instructor? (select all that apply)
▢ ChatGPT
▢ Online Homework Help
▢ Study Groups
▢ Supplemental Instructors
▢ Private Tutoring
▢ Other (please specify)

What uses of ChatGPT do you think would be beneficial for students? (check all that apply)
▢ Personalized Learning
▢ Gamification
▢ Effective and Instant Feedback
▢ Progress Tracking
▢ Adjusting difficulty of material
▢ Other (please specify)

How much do you agree or disagree with this statement: "ChatGPT will enable academic dishonest behaviors."
o Strongly agree
o Somewhat agree
o Neither agree nor disagree
o Somewhat disagree
o Strongly disagree

How do you think ChatGPT will impact:

| | *Extremely negative* | *Somewhat negative* | *Neither positive nor negative* | *Somewhat positive* | *Extremely positive* |
|---|---|---|---|---|---|
| Critical Thinking | | | | | |
| Problem Solving | | | | | |
| Teamwork | | | | | |
| Self Efficacy | | | | | |
| Test Anxiety | | | | | |
| Academic Performance | | | | | |
| Intrinsic Motivation | | | | | |
| Student Engagement | | | | | |

How do you think AI tools like ChatGPT will impact the future of your discipline?

**B. ChatGPT Student Survey**
Dear TAMU students,

A team of researchers led by Dr. Kristi Shryock are interested in understanding how ubiquitous the AI platform ChatGPT is in the lives of our students. This short survey is seeking to gauge the use cases of ChatGPT among students and understand the prominence of ChatGPT among our students at Texas A&M University. All responses to this survey will be anonymous.

Please select the college/school and department/program you are associated with
College/School
Department/Program
▼ College of Agriculture and Life Sciences... Transition Academic Programs ~ General Studies

What is your graduating class?

How familiar are you with ChatGPT by OpenAI?
o Extremely familiar
o Very familiar
o Moderately familiar
o Slightly familiar
o Not familiar at all

Have you made an account and used ChatGPT for any reason (personal or educational)?
o No
o Yes

(If yes to account) When did you decide to create your account? (month or semester you began using ChatGPT)

(If yes to account) In what ways have you used ChatGPT? (select all that apply)
▢ Asking technical questions
▢ Carrying on a conversation out of curiosity
▢ Asking general knowledge questions / advice
▢ Solving homework
▢ Checking solutions
▢ Asking quick questions when stuck on a problem
▢ Explaining concepts
▢ Writing essays
▢ Have not used
▢ Other (Please Specify)

In what ways do your peer students use ChatGPT? (select all that apply)

▢ Asking technical questions
▢ Carrying on a conversation out of curiosity
▢ Asking general knowledge questions / advice
▢ Solving homework
▢ Checking solutions
▢ Asking quick questions when stuck on a problem
▢ Explaining concepts
▢ Writing essays
▢ Have not used
▢ Other (Please Specify)

How much do you agree or disagree with this statement: "Students should be allowed to utilize resources not provided by the instructor of a course." (Ex: Chegg, Coursehero, ChatGPT, Quizlet)

o Strongly agree
o Somewhat agree
o Neither agree nor disagree
o Somewhat disagree
o Strongly disagree

To what extent is using ChatGPT ethical / appropriate for coursework?

o Extremely appropriate
o Somewhat appropriate
o Neither appropriate nor inappropriate
o Somewhat inappropriate
o Extremely inappropriate

How much do you agree or disagree with this statement: "ChatGPT will enable academic dishonest behaviors."

o Strongly agree
o Somewhat agree
o Neither agree nor disagree
o Somewhat disagree
o Strongly disagree

How do you think ChatGPT will impact:

| | *Extremely negative* | *Somewhat negative* | *Neither positive nor negative* | *Somewhat positive* | *Extremely positive* |
|---|---|---|---|---|---|
| Critical Thinking | | | | | |
| Problem Solving | | | | | |
| Teamwork | | | | | |
| Self Efficacy | | | | | |
| Test Anxiety | | | | | |
| Academic Performance | | | | | |
| Intrinsic Motivation | | | | | |

Student Engagement

How do you think AI tools like ChatGPT will impact the future of your discipline?